\documentclass[aps,prd,preprint,preprintnumbers,unsortedaddress,superscriptaddress,showpacs,nofootinbib]{revtex4-1}
\usepackage{graphicx,color}
\usepackage{relsize}
\usepackage{slashed}
\usepackage{color}
\usepackage{ulem}
\usepackage{tabu}

\usepackage{amsmath}
\usepackage{amssymb}
\usepackage{amsthm}
\usepackage{mathrsfs}
\usepackage{graphicx}
\usepackage{epstopdf}
\usepackage{fancyhdr}
\usepackage{array}
\usepackage[all]{xy}
\usepackage{eufrak}
\usepackage{euscript}
\usepackage{enumerate}
\usepackage{slashed}
\usepackage{hyperref}
\usepackage{subfigure} 
\usepackage{epstopdf} 

\usepackage{hyperref}
\hypersetup{pdftex,colorlinks=true,linkcolor=blue,citecolor=blue,menucolor=black,urlcolor=blue,filecolor=blue}
\def\lag{\langle}
\def\rag{\rangle}

\newcommand{\bK}{\bar{K}}

\begin{document}

\title{The radiative decay $D^0\to \bar{K}^{*0}\gamma$ with vector meson dominance}

\author{J.~M.~Dias}
\email{jdias@if.usp.br}
\affiliation{Institute of Modern Physics, Chinese Academy of Sciences, Lanzhou 730000, China
}
 
\affiliation{Departamento de Física Teórica and IFIC, Centro Mixto Universidad de Valencia-CSIC, 
Institutos de Investigac\'ion de Paterna, Aptdo. 22085, 46071 Valencia, Spain.
}
\affiliation{
Instituto de F\'isica, Universidade de S\~ao Paulo, C.P. 66318, 05389-970 S\~ao 
Paulo, SP, Brazil.
}

\author{V.~R.~Debastiani}
\email{vinicius.rodrigues@ific.uv.es}
\affiliation{Institute of Modern Physics, Chinese Academy of Sciences, Lanzhou 730000, China
}
\affiliation{Departamento de Física Teórica and IFIC, Centro Mixto Universidad de Valencia-CSIC, 
Institutos de Investigac\'ion de Paterna, Aptdo. 22085, 46071 Valencia, Spain.
}
\author{Ju-Jun Xie}
\email{xiejujun@impcas.ac.cn}
\affiliation{Institute of Modern Physics, Chinese Academy of Sciences, Lanzhou 730000, China
}

\author{E.~Oset}
\email{Eulogio.Oset@ific.uv.es}
\affiliation{Institute of Modern Physics, Chinese Academy of Sciences, Lanzhou 730000, China
}
\affiliation{Departamento de Física Teórica and IFIC, Centro Mixto Universidad de Valencia-CSIC, 
Institutos de Investigac\'ion de Paterna, Aptdo. 22085, 46071 Valencia, Spain.
}

\preprint{}

\date{\today}

\begin{abstract}
Motivated by the experimental measurements of $D^0$ radiative decay modes we have proposed a model to 
study the $D^0\to \bar{K}^{*0}\gamma$ decay, by establishing a link with $D^0\to \bar{K}^{*0}V$ 
$(V=\rho^0,\, \omega)$ decays through the vector meson dominance hypothesis. In order to do this properly, 
we have used the Lagrangians from the local hidden gauge symmetry approach to account for 
$V\gamma$ conversion. As a result, we have found the branching ratio $\mathcal{B}[D^0\to \bK^{*0}\gamma]
=(1.55 - 3.44)\times 10^{-4}$, which is in fair agreement with the experimental values reported by Belle and 
Babar collaborations.
\end{abstract}


\maketitle

\section{Introduction}

The heavy hadron weak decays have become an important source of information not 
only in the quest for new physics beyond standard model but also to 
understand in a deeper way the hadron dynamics behind those processes. 
For instance, the $B$ meson decays have been experimentally measured by 
LHCb \cite{lhcb:bsdecay}. A large signal was found for the $f_0(980)$ 
resonance in the $B_s\to J/\psi \pi^-\pi^+$ channel, while no peak associated 
with $f_0(500)$ was reported. The same analysis has been done by CDF 
\cite{cdf}, Belle \cite{belle} and D$0$ \cite{d0} collaborations. In contrast, in 
Ref.~\cite{lhcb:b0decay} the $f_0(500)$ was seen in the $B \to J/\psi\pi^-\pi^+$ 
decay mode, while only a small fraction for $f_0(980)$ was observed. 
Theoretically, these results were understood \cite{liang,bayar} through 
the chiral unitary approach, which implements chiral theory with some unitary 
and symmetry techniques \cite{chu1,chu2}. Furthermore, the application of 
chiral unitary techniques to study the $B$ decays into $J/\psi$ and light vector mesons ($V$) 
provided results which gave support to the interpretation of the vector mesons 
within the $q\bar{q}$ quark picture \cite{qqbar1,qqbar2}. On the other hand, concerning 
to $B$ radiative decays, there is little experimental information 
available on their branching ratios \cite{babar:bgamma,lhcb:bgamma}, 
while the theoretical predictions associated with those ratios differ at least by 
$2$ orders of magnitude, requiring more investigation in order to shed light on this issue 
\cite{qcdfact,qcdpert,qcdfactapp}. Regarding this point, the $B$ radiative decays 
were studied in Ref.~\cite{geng}, where accurate results for these ratios were obtained. 
More concretely, since the long-distance effects might be dominant in those decays, 
the authors presented a mechanism where a link between $B\to J/\psi V$ and 
the $B\to J/\psi\gamma$ decay was established by means of the vector meson dominance 
hypothesis (VMD) \cite{vmd}. The implementation of VMD was done using the 
Lagrangians from local hidden gauge symmetry \cite{bando1,bando2,meissner}. 
The results found in \cite{geng} were in a good agreement with the upper limits set for 
the branching ratios aforementioned.

The charm radiative decays are even more dramatic and have been less 
discussed in the literature. The amount of theoretical work follows the same 
line of the experimental counterpart, i. e. the lack of experimental results 
associated with radiative decays of charmed mesons does not motivate many 
theoretical studies since most of them are dedicated to the search for 
new physics beyond the standard model. It turns out that the charm radiative 
decays are completely dominated by long-distance effects and this feature 
makes them not so attractive to new physics practitioners. On the other hand, 
concerning the hadronic systems, this same feature makes these charmed 
radiative decays an interesting issue to investigate the hadron dynamics 
as well as to make predictions to be tested by the experimental facilities. This 
might be a good scenario to test the successful chiral unitary theory and other 
nonperturbative models related to the description of hadron dynamics.  

As mentioned previously, the amount of experimental information for the charm 
radiative decays is scarce. For instance, the first branching ratio measurement 
for $D^0\to \bK^{*0} \gamma$ radiative decay was performed in 2008 by the Babar collaboration 
\cite{babar}, with $\mathcal{B}(D^0\to \bK^{*0}\gamma)=(3.8 \pm 0.20 \pm 0.27) \times 10^{-4}$, 
where the first error is statistical and the second one systematic. Recently, the 
Belle collaboration has also measured that same branching ratio \cite{belle}, obtaining 
a different value $\mathcal{B}(D^0\to \bK^{*0}\gamma)=(4.66 \pm 0.21 \pm 0.21) \times 10^{-4}$ although it is of the 
same order of the one reported by Babar. From the theoretical side, in Ref.~\cite{biswas} 
the authors have used an extension of the standard model approach in order to 
separate the long and short-range contributions for the $\mathcal{B}(D^0\to \bK^{*0}\gamma)$ 
branching ratio. They have estimated a range of values equal to 
$\mathcal{B} = (4.6-18)\times 10^{-5}$, with large error. On the other hand, using a different approach, a value 
$20$ to $80$ times smaller than the previous one was obtained in Ref.~\cite{shen}: 
$\mathcal{B} = (0.22 \pm 0.03 \pm 0.01)\times 10^{-5}$. In both cases, although 
VMD is invoked in those studies, they aim at the search for new 
physics and because of that the authors are more concerned to what happens 
to the short-range contributions from one model to the other. In view of this, in this 
work we adopt a different perspective and look at what happens to the hadron dynamics in these decays, and 
we shall propose a model based on the mechanism of Ref.~\cite{geng} to estimate 
the $D^0\to\bK^{*0}\gamma$ branching ratio. Although the short-range contributions 
play an important role in $B$ meson decays, in some cases, as that shown by the 
authors of Ref.~\cite{geng}, the long-range physics is the main ingredient and may 
help to provide more accurate results, as it was discussed in that work. Since in the 
charm sector the radiative decays are largely dominated by the long-range physics \cite{belle,biswas,shen}, 
we expect to get reasonably accurate results.

The starting point in our approach is to establish a link between the 
$D^0\to\bK^{*0}V$ decays, with the vector meson $V$ related to the $\rho$ and 
$\omega$ mesons, and the radiative $D^0\to\bK^{*0}\gamma$ decay via VMD hypothesis. 
In our case, the VMD is implemented using the hidden gauge Lagrangians 
\cite{bando1,bando2,meissner}, describing the $V\gamma$ conversion. In the 
next section, we show the details on how to do this properly and 
also how to get the branching ratios we are concerned with. We also present arguments 
that support the suppression of the short-range effects in the amplitudes contributing 
to the branching ratio we are interested in.

\section{Theoretical Framework}

In order to calculate the radiative decay $D^0\to \bar{K}^{*0}\gamma$ 
rate, we follow the approach used in Ref.~\cite{geng}, 
where the authors combine vector meson dominance, through 
hidden gauge Lagrangians, with a novel mechanism, proposed in 
Ref.~\cite{bayar} for $B^0(B^0_s)\to J/\psi V$, to describe the 
$B^0(B_s^{0})\to J/\psi\gamma$ decays. In the following, we shall 
describe briefly this mechanism extended to our problem. 

\begin{figure}
	\begin{center} 
		\includegraphics[width=0.4\textwidth]{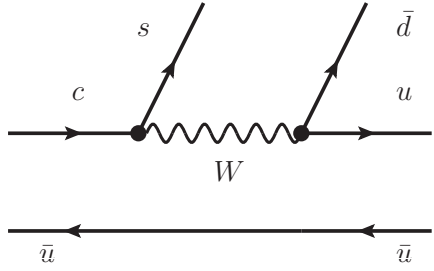}
	\end{center} 
	\caption{\label{diagdkg} The $D^0$ meson decaying weakly into a 
	$\bar{K}^{*0}$ and a vector meson at the quark level: a 
	$c$ quark converts into a $s$ quark by emission of a $W$ boson, 
	which then coalesces into a $\bar{d}u$ pair, producing a $u\bar{u}$ pair 
	in the final state. The first pair of quark/antiquarks forms a $\bar{K}^{*0}$ meson, while 
	the remaining $u\bar{u}$ can either form a $\rho$ or $\omega$ meson.}
\end{figure}

The $D^0$ meson decays weakly into $\bar{K}^{*0}$ meson in addition 
to a $\rho^0$ or $\omega$ meson, denoted by $V$. At the 
quark level, this process is illustrated in Fig.~\ref{diagdkg}. According to this 
figure, a $c$ quark converts into a strange quark by emission of 
a $W$ boson, that subsequently coalesces into a $\bar{d}u$ pair. As 
a result, we have a $\bar{K}^{*0}$ meson, related to the $s\bar{d}$ pair, 
while the remaining $u\bar{u}$ can be related to the vector mesons, 
$\rho^0$ or $\omega$. It is worth to emphasize at this point that we 
adhere the $q\bar{q}$ picture for vector mesons. In fact, studies have 
shown that wave functions for the low-lying vector mesons are essentially 
dominated by $q\bar{q}$ components 
\cite{qqbar1,weinberg,hanhart,gamermann,hyodo,sekihara,aceti,xiao}. 
Therefore, in terms of quarks the wave functions for vector mesons are given 
by \footnote{In general, the physical isoscalars $\phi$ and $\omega$ are mixtures 
of the SU(3) wave functions $\psi_8$ and $\psi_1$:
\begin{eqnarray}
\phi    &=& \psi_8 {\rm cos}\theta - \psi_1 {\rm sin}\theta \, , \nonumber\\
\omega  &=& \psi_8 {\rm sin}\theta + \psi_1 {\rm cos}\theta \, ,\nonumber
\end{eqnarray}
where $\theta$ is the nonet mixing angle and:
\begin{eqnarray}
\psi_8 &=& \frac{1}{\sqrt{6}} (u\bar{u} + d \bar{d} -2 s \bar{s}) \,\nonumber
, \\
\psi_1 &=& \frac{1}{\sqrt{3}} (u\bar{u} + d \bar{d} + s \bar{s}) \,.\nonumber
\end{eqnarray}
For ideal mixing, ${\rm tan}\theta = 1/\sqrt{2}$ (or $\theta =
35.3^0$), the $\omega$ meson is pure $u\bar{u} + d\bar{d}$, and the $\phi$ 
meson becomes pure $s\bar{s}$ state.}
\begin{eqnarray}
\rho^0 &=& \frac{1}{\sqrt{2}}(u\bar{u}-d\bar{d})\, ,\nonumber\\
\omega &=& \frac{1}{\sqrt{2}} (u\bar{u}+d\bar{d})\, \nonumber.
\end{eqnarray}
Since there is no $s\bar{s}$ pair in the process of Fig.~\ref{diagdkg}, 
we do not have $\phi$ meson contribution.

In order to write the $D^0 \to \bar{K}^{*0} V$ amplitudes we restrain ourselves to 
factorize the weak vertices in terms of a factor $V^{\prime}_p$, which contains 
weak vertices, Cabibbo angles, etc. The factor $V^{\prime}_p$ gets canceled 
since we are interested in ratios of decay rates. A similar assumption was done in 
Ref.~\cite{bayar}, where the decay rates related to 
$\bar{B}^0\to J/\psi \bar{K}^{*0}$ and $\bar{B}^0_s\to J/\psi K^{*0}$ 
channels were evaluated, with results in good agreement with the 
experimental ones \cite{pdg}. Hence, the amplitudes for the $\bar{K}^{*0}V$ 
production are
\begin{eqnarray}\label{tdkg}
t_{D^0\to \bar{K}^{*0} \rho^0} &=& \frac{V^{\prime}_p}{\sqrt{2}}\, ,\nonumber\\
t_{D^0\to \bar{K}^{*0} \omega} &=& \frac{V^{\prime}_p}{\sqrt{2}}\, ,
\end{eqnarray}
where polarization vectors in each expression above are omitted (we shall 
come back later on about the spin structure).

Once we have determined the amplitudes associated with the production 
of $\bar{K}^{*0}V$, we have to go a step further and let the $V$ meson 
convert into a photon $\gamma$, according to VMD-hypothesis \cite{vmd}. 
In order to implement VMD properly, we use the Lagrangians from the Local hidden 
gauge approach \cite{bando1,bando2,meissner}, which for the $V\gamma$ vertex 
are given by
\begin{eqnarray}\label{lhg}
\mathcal{L}_{V\gamma} = - M^2_V \frac{e}{g} A_{\mu} \lag V^{\mu}Q \rag \, ,
\end{eqnarray}
where $e$ is the electron charge, $e^2/4\pi \approx 1/137$, and $g$ is the 
universal coupling in the hidden gauge Lagrangian, defined by $g= M_V/(2f_{\pi})$, 
with $f_{\pi}$ the pion decay constant ($f_{\pi}=93$ MeV), while $M_V$ stands for the vector 
meson mass (we take $M_V=780$ MeV in this work). $A_{\mu}$ 
is associated with the photon field and $V^{\mu}$ is the matrix below
\begin{equation}
\label{Vmatrix}
V_{\mu} =
\left(
\begin{array}{ccc}
\frac{1}{\sqrt{2}} \rho^0 + \frac{1}{\sqrt{2}} \omega & \rho^+ & K^{* +} \\
\rho^- & -\frac{1}{\sqrt{2}} \rho^0 + \frac{1}{\sqrt{2}} \omega & K^{* 0} \\
K^{* -} & \bar{K}^{* 0} & \phi \\
\end{array}
\right)_{\mu}\, .
\end{equation}
Furthermore, in Eq.~\eqref{lhg} $Q=\textrm{diag}(2/3,-1/3,-1/3)$ is the 
charge matrix of the $u,\,d$, and $s$ quarks, while the symbol 
$\lag\,\,\, \rag$ in Eq.~\eqref{lhg} means the trace of the $V^{\mu}Q$ product. Therefore 
the Lagrangian of Eq.~\eqref{lhg} now reads
\begin{equation}\label{lvg}
\mathcal{L}_{V\gamma} = - M^2_V \frac{e}{g} A_{\mu} \Big[ \frac{\rho^{\mu}}{\sqrt{2}}+ 
\frac{\omega^{\mu}}{3\sqrt{2}} - \frac{\phi^{\mu}}{3}\Big] \, .
\end{equation}
Eq.~\eqref{lvg} can be simplified if we define 
$\tilde{V}^{\mu}$ as denoting the $\rho^0$, $\omega$ and $\phi$ fields 
and $C_{V\gamma}$ standing for their respective constants 
$1/\sqrt{2},\,1/3\sqrt{2},\, -1/3$. Therefore, we have
\begin{equation}\label{lvgamma}
\mathcal{L}_{V\gamma} = - M^2_V \frac{e}{g} A_{\mu} \tilde{V}^{\mu}\, C_{V\gamma} \, ,
\end{equation}
with $C_{V\gamma}$ given by
\begin{equation}\label{const}
C_{\gamma V}=\left\{\begin{array}{ll}
\frac{1}{\sqrt{2}} & \mbox{for $\rho^0$}\\
\frac{1}{3}\frac{1}{\sqrt{2}}&\mbox{for $\omega$}\\
-\frac{1}{3} &\mbox{for $\phi$}\end{array}
\right. .
\end{equation}

\begin{figure}
	\begin{center} 
		\includegraphics[width=0.4\textwidth]{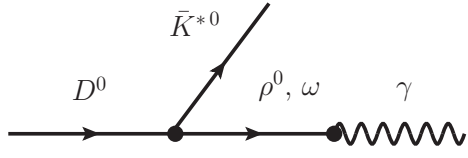}
	\end{center} 
	\caption{\label{diagdkghalevel} Diagram associated with $D^0\to\bar{K}^{*0}\gamma$ 
	decay in the hadron level.}
\end{figure}

Now that we have determined the $\bar{K}^{*0}V$ production 
amplitude as well as the Lagrangian that describes the 
$V\gamma$ vertex, we can write down the amplitude for the 
photon production, which is depicted in Fig.~\ref{diagdkghalevel}.

Using Eqs.~\eqref{tdkg}, \eqref{lvgamma} and \eqref{const}, the 
decay amplitude for the diagram of Fig.~\ref{diagdkghalevel} is
\begin{eqnarray}\label{tgammaprod}
-it_{D^0\to \bar{K}^{0}*\gamma}&=&-it_{D^0\to \bar{K}^{*0}\rho^0}\, \epsilon_{\mu}(\rho)
\frac{i}{q^2-M^2_{\rho}}(-i) M^2_{\rho} \frac{e}{g}\,\epsilon_{\nu}(\rho)\, \epsilon^{\nu}(\gamma)
\,C_{\rho\gamma}\nonumber\\
&-&it_{D^0\to \bar{K}^{*0}\omega}\, \epsilon_{\alpha}(\omega)\,
\frac{i}{q^2-M^2_{\omega}}(-i) M^2_{\omega} \frac{e}{g}\,\epsilon_{\beta}(\omega)\, 
\epsilon^{\beta}(\gamma)\,C_{\omega \gamma}\, ,
\end{eqnarray}
where $\epsilon_{\mu}(\rho)$, $\epsilon_{\alpha}(\omega)$ are 
the polarization vectors for the $\rho^0$ and $\omega$ mesons production, 
while $\epsilon^{\nu}(\gamma)$ is the one associated with the photon. 
Remembering that
\begin{equation}
\sum\limits_{\textrm{pol.}} \epsilon_{\mu}(V)\,\epsilon_{\nu}(V) = -g_{\mu\nu}
 + \frac{p^V_{\mu}p^V_{\nu}}{M^2_V}\, ,
\end{equation}
and knowing that $p^V\cdot \epsilon(V)=0$ (Lorentz condition), with 
$p^V_{\mu}$ the momentum of the vector meson 
that is equal to the photon one, after a bit of algebra 
Eq.~\eqref{tgammaprod} can be written as 
\begin{equation}\label{gprod}
t_{D^0\to \bar{K}^{*0}\gamma}=\frac{2}{3}\frac{e}{g}\,V^{\prime}_p\,\epsilon_{\mu}(\gamma)\, ,
\end{equation}
where we have used the approximation $M_{\rho}\approx M_{\omega}\approx M_V$, as it is 
usual in the hidden gauge approach.

In order to estimate the ratios we need the decay formulas 
associated with the $D^0\to \bar{K}^{*0}\rho^{0}\,(\omega)$ and 
$D^0\to \bar{K}^{*0}\gamma$ channels. They are given by
\begin{eqnarray}\label{decay}
\Gamma_{D^0\to \bar{K}^*\rho(\omega)}&=&\frac{1}{8\pi}\frac{1}{M^2_{D}}
\sum_{\textrm{pol.}} \Big| t_{D^0\to \bar{K}^{*0}\rho^0(\omega)} \Big|^2 p_{\rho(\omega)}\, ,\nonumber\\
\Gamma_{D^0\to \bar{K}^{*0}\gamma}&=&\frac{1}{8\pi}\frac{1}{M^2_{D}}\,
\sum_{\textrm{pol.}}\Big|t_{D^0\to \bar{K}^{*0}\gamma}\Big|^2 p_{\gamma}\, ,
\end{eqnarray} 
where $p_{\rho(\omega)}$ and $p_{\gamma}$ are the $\rho^0(\omega)$ meson 
and the photon momenta in the $D^0$ rest frame. Using 
Eqs.~\eqref{tdkg} and \eqref{gprod} into Eq.~\eqref{decay}, we get the 
following expression for the ratio $\Gamma_{D\to\bar{K}^{*0}\gamma}
/\Gamma_{D^0\to\bar{K}^{*0}\rho^0}$
\begin{equation}\label{ratio}
\frac{\Gamma_{D^0\to\bar{K}^{*0}\gamma}}{\Gamma_{D^0\to\bar{K}^{*0}\rho^0}}=\Big(\,\frac{2}
{3}\frac{e}{g}\, 
\Big)^2 \,\frac{p_{\gamma}}{p_{\rho}}\, .
\end{equation}
As we mentioned before, the parametrization of the weak vertex 
defined as $V^{\prime}_p$ does not play a role in our 
approach since it gets canceled, as can be seen by looking at the ratio 
in Eq.~\eqref{ratio}.

\begin{figure}
	\begin{center} 
		\includegraphics[width=1.0\textwidth]{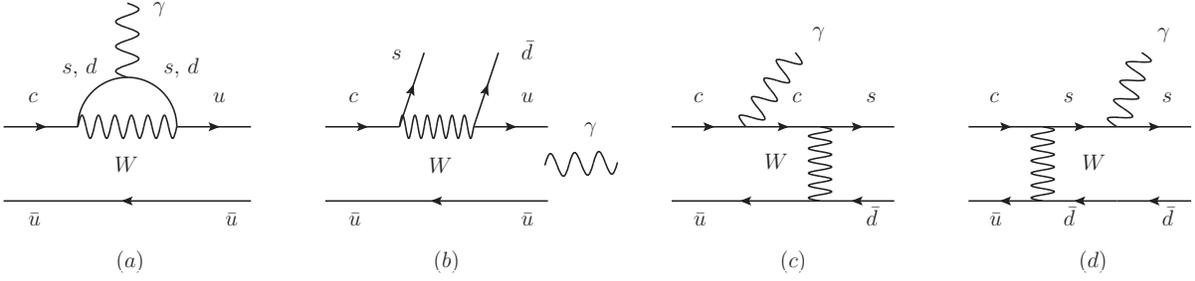}
	\end{center} 
	\caption{\label{slrange} Short- and long-range diagrams contributing to $D^0\to\bar{K}^{*0}\gamma$ amplitude.}
\end{figure}

In a general context the mechanism that we have adopted here is considered 
as a long range process in Refs.~\cite{cheng,golowich,golowich2,donoghue,burdman}. 
In these works, the $B$ radiative decays involving a $K^*$ and $\rho$ mesons 
were addressed. They were separated into long and short range processes and their 
contribution was estimated. As a result, the short range 
contribution, considered in those works as the dominant one, for the $B\to K^*\gamma$ 
process was bigger (by a factor $30$) than the upper bounds for the 
$B\to J/\psi\gamma$ case, indicating that the equivalent short range contribution 
could not be dominant in the $J/\psi\gamma$ case, as discussed in Ref.~\cite{geng}. 
Furthermore, in the charm sector, it was pointed out in Ref.~\cite{burdman} that 
the short range diagrams provided results smaller than the one related to the its 
long range counterpart. In our case the short range diagram gives no contribution 
since there is no $\bar{K}^{*0}(s\bar{d})$ production in the final state, as can be seen in 
Fig.~\ref{slrange}(a). 

In Fig.~\ref{slrange} we show all the diagrams associated with short and long 
range processes. As we have mentioned previously, the diagram of 
Fig.~\ref{slrange}(a) related to the short range contribution, 
does not contribute in our case, which is represented by the diagram (b), since it 
produces $\rho^0 \gamma$ or $\omega \gamma$ but not $\bK^{*0} \gamma$. 
The remaining ones, Fig.~\ref{slrange}(c)-(d), are suppressed with respect to that 
in Fig.~\ref{slrange}(b). This happens, because they have 
a weak process involving two quarks of the original $D^0$ meson and, according 
to the discussion in Ref.~\cite{xie}, this kind of processes are penalized with 
respect to those involving just one quark.

According to Ref.~\cite{geng} we have to take into account the 
polarization structure of the $D^0\to \bar{K}^{*0}\gamma$ vertex. 
In weak decay processes we can have parity violation as well as 
parity conservation. In order to take this feature into account in 
our model, we are going to follow the procedure of Ref.~\cite{geng} 
and define both parity conserving ($PC$) and parity violating ($PV$) 
structures, which are often used in weak decay studies \cite{qcdfact,qcdpert,qcdfactapp,golowich2}. They are 
\begin{equation}\label{pc}
V_{PC}=\frac{V^{\prime}_p}{\sqrt{2}}\,\epsilon_{\mu\nu\alpha\beta}\,\epsilon^{\mu}(\bar{K}^*)q^{\nu}
\epsilon^{\prime\, \alpha}(V)q^{\prime\, \beta}\, ,
\end{equation}
\begin{equation}\label{vp}
V_{PV}=\frac{V^{\prime}_p}{\sqrt{2}}\,\epsilon_{\mu}(\bar{K}^*)\epsilon^{\prime\, \nu}(V)
(g_{\mu\nu}\,q\cdot q^{\prime} - q_{\mu}^{\prime}q_{\nu})\, ,
\end{equation}
where $\epsilon^{\prime}(V)$ and $q^{\prime}$ are the 
polarization vector as well as the momentum of the vector 
meson ($\rho^0$ or $\omega$) to be converted into $\gamma$ 
through VMD. In the case of photon production in both 
Eqs.~\eqref{pc} and \eqref{vp}, $\epsilon^{\prime}$ as well as 
$q^{\prime}$ stand for the vector polarization and momentum 
of the photon, respectively.

Note that both structures are gauge invariant. In fact, when 
we use the Lagrangians from local hidden gauge approach 
to deal with vector-vector interactions $VV$ and also $V\gamma$ 
conversion, gauge invariant amplitudes are obtained, as discussed in 
Refs.~\cite{bando1,bando2,meissner,nagahiro,nagahiro2}. 

In order to take into account the polarization structure of the 
weak vertices, as discussed previously, we have to sum the 
Eqs.~\eqref{pc} and \eqref{vp} over the polarizations of the 
vector meson or the photon. Summing up over the polarization 
provides
\begin{eqnarray}
\sum\limits_{\lambda}\sum\limits_{\lambda^{\prime}} |V_{PC}|^2&=&2[(q\cdot q^{\prime})^2 - q^2q^{\prime\,2}]\, ,\nonumber\\
\sum\limits_{\lambda}\sum\limits_{\lambda^{\prime}} |V_{PV}|^2&=&2(q\cdot q^{\prime})^2 + q^2q^{\prime\,2}\, ,
\end{eqnarray}
where $q^{\prime\, 2}=M^2_V$ for vector production or $0$ in the case of photon 
production, while
\begin{equation}
q\cdot q^{\prime} = \frac{1}{2}(M^2_{D^0} - M^2_{\bar{K}^{*0}} - M^2_{V})\, ,
\end{equation}
with $M_V^2=0$ for the case of photon production.

With this, we can obtain the following factors
\begin{equation}\label{fpc}
R_{PC}=\frac{\sum\limits_{\lambda}\sum\limits_{\lambda^{\prime}} |V_{PC}|^2\, \textrm{for\,}\gamma}
{\sum\limits_{\lambda}\sum\limits_{\lambda^{\prime}} |V_{PC}|^2\, \textrm{for\,}\rho}\, ,
\end{equation}
and
\begin{equation}\label{fpv}
R_{PV}=\frac{\sum\limits_{\lambda}\sum\limits_{\lambda^{\prime}} |V_{PV}|^2\, \textrm{for\,}\gamma}
{\sum\limits_{\lambda}\sum\limits_{\lambda^{\prime}} |V_{PV}|^2\, \textrm{for\,}\rho}\, .
\end{equation}

Therefore, the polarization structures discussed above are 
taken into account in our calculation simply by plugging them 
in Eq.~\eqref{ratio}, which now reads
\begin{equation}\label{bf}
\frac{\mathcal{B}(D^0\to \bK^{*0} \gamma)}
{\mathcal{B}(D^0\to \bK^{*0}\rho)}=\Big(\,\frac{2}
{3}\frac{e}{g}\, \Big)^2 \,\Big(\,\frac{p_{\gamma}}{p_{\rho}}\,\Big)\, R_{PC(PV)}\, ,
\end{equation}
where in the left-hand side we have divided the numerator as well as the denominator 
by $\Gamma_{total}$ in order to convert the widths into branching fractions.

\section{Results}

In order to estimate our results, we use the following values for the 
meson masses: $M_{\rho}\approx M_{\omega}\approx M_V =780$ MeV, $M_{\bar{K}^*}=891.6$ 
MeV and $M_{D^0}=1864.8$ MeV. Furthermore, we also 
use as an input for $\Gamma_{D^0\to\bar{K}^*\rho\,(\omega)}$ an average value from 
the following experimental results, extracted from PDG \cite{pdg}, which in our approach 
should be equal. We have
\begin{eqnarray}\label{expratio}
\mathcal{B}[D^0\to\bar{K}^{*0}\rho^0] &=& (1.58 \pm 0.35) \times 10^{-2}\, ,\nonumber\\
\mathcal{B}[D^0\to\bar{K}^{*0}\omega] &=& (1.10 \pm 0.5) \times 10^{-2}\, .
\end{eqnarray}
These results are compatible, within errors, providing an average value of 
$(1.4 \pm 0.4 ) \times 10^{-2}$. Therefore, from Eq.~\eqref{bf}, using the 
values defined above, we get the following result for the branching fraction associated with 
$D^0\to \bK^{*0}\gamma$ channel
\begin{equation}\label{theoratio1}
\mathcal{B}[D^0\to \bar{K}^{*0}\gamma] = (3.44 \pm 1.0)\times 10^{-4},\, \textrm{for PV}
\end{equation}
\begin{equation}\label{theoratio2}
\mathcal{B}[D^0\to \bar{K}^{*0}\gamma] = (1.60 \pm 0.5)\times 10^{-4},\, \textrm{for PC}\, ,
\end{equation}
where the uncertainties are obtained from the experimental errors. The average experimental 
value in the PDG \cite{pdg} is
\begin{equation}
\mathcal{B}[D^0 \to \bK^{*0}\gamma]_{\textrm{exp}}=(4.1 \pm 0.7) \times 10^{-4}\, .
\end{equation}
We can see that the theoretical result with $PV$ is compatible with the experimental number. The one 
with $PC$ is somewhat smaller. An equal mixture of both the $PC$ and $PV$ modes would 
give
\begin{equation}\label{average}
\mathcal{B}[D^0 \to \bK^{*0}\gamma]=(2.5\pm 1.1)\times 10^{-4}\, ,
\end{equation}
which is compatible with the experimental value within errors.

In Ref.~\cite{biswas} the authors have used a model related to the extensions of the 
standard model in order to look for new physics in the charm rare decays. They 
have done calculations for the long range distance $D\to V\gamma$ 
amplitudes (see Table IV of that reference), where for the $D^0\to\bar{K}^{*0}\gamma$ a ratio 
of about $(4.6 - 18)\times 10^{-5}$ was obtained. Using a different approach called 
light-front quark model, a similar result was found in Ref.~\cite{shen}. The value 
obtained in this latter work for the same ratio was $(4.5 - 19)\times 10^{-5}$. Note that 
both results are smaller than our result, given by 
Eqs.~\eqref{theoratio1}, \eqref{theoratio2} and \eqref{average}, as well as the experimental one. Note also that the range 
of allowed values is much bigger than in our case, and the lower bound is about one order 
of magnitude smaller than our results.

\section{Conclusions}

Using a mechanism adopted in Ref.~\cite{geng} we have established 
a link between the $D^0\to \bK^{*0} V$, with $V=\rho^0,\, \omega$ mesons, 
and $D^0\to \bK^{*0}\gamma$ radiative decays. Concretely, after 
calculating the amplitude for $V$ meson production, we use the 
vector meson dominance hypothesis in order to convert the vector mesons 
produced in our mechanism into a photon. This was done using the 
Lagrangians from the local hidden gauge approach, which provides a 
gauge invariant amplitude when the vector polarization structure is 
taken into account. Thus, we have obtained an expression in which both 
parity violation and conservation contributions are considered. As a result, 
we have obtained a value for the branching ratio $\mathcal{B}[D^0\to \bK^{*0} 
\gamma]$ in a fairly agreement with the experimental value quoted in 
the PDG \cite{pdg}, while other estimations using different approaches provide results with 
large uncertainties, with some values one order of magnitude smaller than our findings.

We should mention that our evaluation is done using as input the experimental rates 
for $D^0\to \bK^{*0}\rho^0 (\omega)$. Alternative calculations that use other experimental 
information to fix unknown parameters of the theory \cite{biswas,shen} lead to 
larger uncertainties. Note that other terms, like loop corrections, 
that in other approaches must be calculated explicitly, are incorporated 
empirically in our approach when using the empirical values of the 
$D^0\to \bK^{*0}\rho^0 (\omega)$ rates \cite{geng}. In this sense, once one 
shows that short range terms in the process studied do not contribute, or are 
small, the method used here proves to be rather accurate for evaluating 
this kind of radiative decays.

\section*{Acknowledgments}

J.~M.~Dias would like to thank the Brazilian funding agency FAPESP for the financial support 
under Grant No. $2016/22561{\rm-}2$.
V.~R.~Debastiani wishes to acknowledge the support from the
Programa Santiago Grisolia of Generalitat Valenciana (Exp. GRISOLIA/2015/005).
This work is also partly supported by the Spanish Ministerio de Economia
y Competitividad and European FEDER funds under the contract number
FIS2014-57026-REDT, FIS2014-51948-C2-1-P, and FIS2014-51948-C2-2-P, and
the Generalitat Valenciana in the program Prometeo II-2014/068. 
This work is partly supported by the National Natural Science Foundation
of China (Grants No. 11475227 and 11735003) and the
Youth Innovation Promotion Association CAS (No. 2016367).

\end{document}